\documentclass{article}
\usepackage{graphicx}
\usepackage{lipsum} 
\usepackage{abstract}
\usepackage{url}
 
\usepackage{geometry}
\usepackage{booktabs} 
\usepackage{cleveref} 
\usepackage{float}
\usepackage{tikz}
\usetikzlibrary{shapes.geometric, arrows, positioning}

\tikzstyle{block} = [rectangle, rounded corners, minimum width=2.8cm, minimum height=0.9cm,text centered, draw=black, fill=blue!10]
\tikzstyle{arrow} = [thick,->,>=stealth]
\tikzstyle{layer} = [rectangle, rounded corners, minimum width=10m, minimum height=1cm,text centered, draw=black, fill=gray!15]

\title{HSM and TPM Failures in Cloud: \\A Real-World Taxonomy and Emerging Defenses}

\author{
  Shams Shaikh\thanks{*Corresponding author. Email: shamsxshaikh@gmail.com (Shams Shaikh)}\\
  Department of Electronics and Computer Science Engineering\\
  Don Bosco College of Engineering, Goa University\\
  Fatorda, Goa, India\\
  \texttt{shamsxshaikh@gmail.com}
  \and
  Trima P. Fernandes e Fizardo\\
  Assistant Professor\\
  Department of Electronics and Computer Science Engineering\\
  Don Bosco College of Engineering, Goa University\\
  Fatorda, Goa, India\\
  \texttt{trima.fernandes@dbcegoa.ac.in}
}

\begin{document}
\date{} 

\maketitle

\begin{abstract}
As cloud infrastructure becomes the backbone of modern organizations, the security of cryptographic key management, especially using Hardware Security Modules (HSMs) and Trusted Platform Modules (TPMs) faces unprecedented challenges. While these hardware-based solutions offer strong protection in isolated environments, their effectiveness is being undermined by cloud-native threats such as misconfigurations, compromised APIs, and lateral privilege escalations.

This paper presents a comprehensive analysis of publicly disclosed attacks and breaches involving HSMs and TPMs in cloud environments, identifying recurring architectural and operational flaws. We propose a taxonomy of attack vectors based on real-world case studies and threat intelligence reports, highlighting the gaps between hardware trust anchors and dynamic cloud ecosystems.

Furthermore, we evaluate emerging defensive paradigms: confidential computing, post-quantum cryptography, and decentralized key management systems (dKMS)—assessing their potential to address these gaps. Our findings emphasize that securing cloud-based cryptographic trust requires a layered, context-aware approach that integrates both hardware and software safeguards. The study serves as a practical framework for cloud architects and security engineers to reassess key protection strategies in light of evolving threats.

To our knowledge, this is the first work to synthesize documented, real-world cloud HSM and TPM failures into a coherent taxonomy grounded in modern threat models.
\end{abstract}

\vspace{1em}

    \vspace{2em}

\section{Introduction}
The security of cryptographic keys is the backbone of modern cybersecurity\cite{legit2025}. Organizations rely on encryption to protect sensitive data, ensure secure transactions, and maintain digital trust. However, the effectiveness of encryption entirely depends on the security of the cryptographic keys themselves. If attackers gain access to these keys, even the strongest encryption becomes useless.

To mitigate this risk, HSMs\cite{enm2024} and TPMs were introduced as trusted hardware solutions to generate, store, and manage cryptographic keys in a highly secure manner. HSMs are dedicated hardware appliances designed to safeguard encryption keys from external threats\cite{utimaco2023}, while TPMs provide built-in cryptographic functions at the hardware level within computing devices. These technologies have been widely adopted in cloud infrastructures, where they are used to secure data at rest, encrypt communication channels, and authenticate critical transactions\cite{fortanix2025}.

However, the rise of cloud computing has introduced new security challenges that traditional HSMs and TPMs were never designed to address\cite{usenix2015}. 

\subsection{Why Are HSMs and TPMs Failing in Cloud Security?}
In traditional on-premise environments, organizations had full control over their hardware security modules. They could physically restrict access, enforce strict security policies, and maintain an air-gapped infrastructure if needed. Cloud environments, however, introduce a completely different threat model - one that traditional HSM and TPM architectures were never designed to handle.

Unlike on-premise deployments, cloud-based HSMs and TPMs face unique challenges, including:

\begin{itemize}
    \item \textbf{API-driven attacks\cite{akamai2024}} – Cloud-based HSMs expose APIs for remote management and key operations. Attackers have exploited weak API authentication and misconfigured permissions to extract encryption keys remotely.
    \item \textbf{Privilege escalation vulnerabilities} – Misconfigured roles and permissions in cloud environments have allowed attackers to gain administrative access to HSMs and TPMs, bypassing security controls entirely.
    \item \textbf{Multi-tenancy risks} – Cloud providers host multiple clients on shared infrastructure. A vulnerability in one tenant’s HSM instance can potentially expose encryption keys to other tenants.\cite{usenix2015}
\end{itemize}

These challenges have led to multiple high-profile security incidents. For example, in 2023, a major cloud provider suffered a security breach where attackers exploited API vulnerabilities in a cloud-based HSM implementation, allowing them to extract sensitive cryptographic keys. This breach not only compromised encrypted data but also undermined trust in the cloud provider’s security framework.

\subsection{Scope of This Research}
This paper aims to critically analyze the shortcomings of HSMs and TPMs in cloud environments by:

\begin{itemize}
    \item Examining real-world attacks that have exposed weaknesses in cloud-based HSM and TPM implementations\cite{wiz2023}\cite{wiz2021chaosdb}.
    \item Identifying key vulnerabilities such as misconfigurations, API risks, and privilege escalation flaws\cite{akamai2024}.
    \item Exploring emerging security alternatives like confidential computing\cite{azure2024confcomp}, post-quantum cryptography\cite{nistpqc2024}, and decentralized key management\cite{fireblocksmpc2023} to determine whether they provide a more effective approach to cryptographic security in the cloud.
\end{itemize}

\subsection{Key Research Questions}
To address these challenges, this research seeks to answer the following critical questions:

\begin{itemize}
    \item Why do HSMs and TPMs fail in cloud environments?
    \item What real-world attacks have exposed their weaknesses?
    \item Are there better alternatives for cloud-based cryptographic key management?
\end{itemize}

By answering these questions, this paper provides actionable insights for cloud architects, security professionals, and organizations seeking to enhance their cryptographic security posture. Ultimately, this research contributes to the ongoing evolution of cloud security strategies, ensuring that encryption remains a trusted safeguard rather than a single point of failure.

\section{Background and Security Failures of HSMs and TPMs in Cloud Environments}

\subsection{The Role of HSMs and TPMs in Cloud Security}
Hardware Security Modules (HSMs) and Trusted Platform Modules (TPMs) serve as specialized hardware guardians for cryptographic operations\cite{hosseinzadeh2020}. HSMs provide isolated environments for encryption, decryption, and key storage, while TPMs establish hardware-based trust for cloud identity systems. Major cloud providers implement these through:
\begin{itemize}
    \item AWS CloudHSM
    \item Azure Key Vault (HSM-backed)
    \item Google Cloud HSM
    \item Virtualized TPMs for cloud instances
\end{itemize}
These modules are often treated as trust anchors, but their effectiveness in cloud-native architectures is increasingly under scrutiny. 
\subsection{How Cloud Breaks Traditional Security Models}
While HSMs and TPMs were designed for tightly controlled on-premise infrastructures, migrating these technologies to the cloud introduces a radically different threat model\cite{usenix2015}. On-premise security relies on physical control, but cloud environments introduce critical vulnerabilities:
\begin{itemize}
    \item Third-party trust dependencies
    \item API-driven attack surfaces
    \item Permission management complexity
    \item Loss of physical isolation
\end{itemize}

\subsection{Real-World Security Failures}
The following case studies illustrate how cloud-native weaknesses, rather than flaws in cryptographic algorithms, have led to major security breaches involving HSMs, TPMs, or key management systems.

\subsubsection{Case Study: Capital One AWS Breach (2019)}

One of the most widely reported examples of IAM misconfiguration compromising cloud cryptographic security is the 2019 Capital One breach\cite{sploit2019}. The attacker, Paige Thompson, exploited a vulnerability in Capital One’s Web Application Firewall (WAF) configuration, using a Server-Side Request Forgery (SSRF) attack to query the AWS EC2 instance metadata service\cite{krebson2019}. This allowed her to retrieve temporary IAM credentials associated with the instance’s assigned role.

These credentials provided access not only to Amazon S3 buckets containing sensitive data, but also potentially to AWS Key Management Service (KMS) and CloudHSM operations, depending on the scope of the attached permissions. If decryption or key export privileges were included, the attacker could have used legitimate channels to access cryptographic keys or decrypted data without bypassing encryption algorithms\cite{techtarget2019}.

This incident illustrates a critical point: the strength of HSMs and cryptographic modules is irrelevant if the surrounding identity and access management (IAM) framework is poorly configured. Effective key security in cloud environments depends not only on the hardware but also on the enforcement of least-privilege access controls and proper segmentation of key management operations\cite{khan2022systematic}.

\subsubsection{Case Study: Azure Cosmos DB ChaosDB Vulnerability (2021)}

In August 2021, security researchers at Wiz disclosed a critical vulnerability in Azure Cosmos DB, dubbed \textit{ChaosDB}, which demonstrated how cloud misconfigurations can undermine cryptographic key protections without compromising encryption algorithms directly \cite{wiz2021chaosdb}.

The vulnerability originated in the Jupyter Notebook feature, which was enabled by default for new Cosmos DB users starting in February 2021. Through a Server-Side Request Forgery (SSRF) attack and privilege escalation within the notebook container, researchers were able to access the underlying Azure infrastructure and retrieve access tokens and certificates intended for internal service use \cite{msrc2021cosmos}.

This allowed full administrative access to Cosmos DB instances across regions, including the ability to extract primary read-write keys via internal management APIs, effectively bypassing any logical boundaries between customer accounts. Although there was no evidence of exploitation beyond the researchers' proof-of-concept, Microsoft disabled the notebook feature globally and advised customers to regenerate their primary keys immediately.

The ChaosDB incident reinforces that even when data is encrypted, access to cryptographic keys or key-granting credentials, such as those stored or handled within managed services like KMS or HSM-backed roles, can render encryption moot. It highlights the importance of defense-in-depth strategies, including network isolation (e.g., Private Link), least-privilege IAM configurations, continuous key rotation, and audit logging \cite{kibet2021cosmos}.

This case underscores the fragility of trust boundaries in multi-tenant cloud platforms and the necessity of applying zero-trust principles even to first-party service integrations.

\subsubsection{Attack Chain Visualization}
The attack path exploited a default-enabled feature, progressing from SSRF to credential theft to full database compromise -highlighted below:
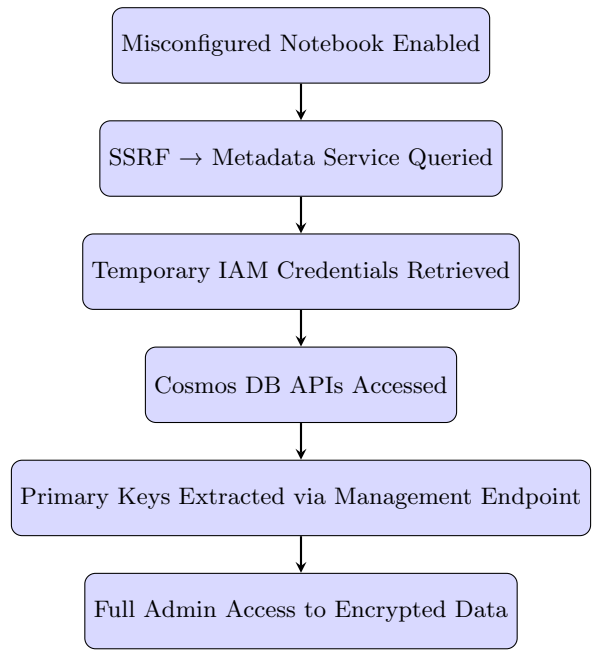
\begin{figure}[h]
\centering
\begin{tikzpicture}[node distance=1.5cm, every node/.style={align=center}, font=\small]
\tikzstyle{startstop} = [rectangle, rounded corners, minimum width=3.8cm, minimum height=1cm,text centered, draw=black, fill=blue!15]
\tikzstyle{arrow} = [thick,->,>=stealth]

\node (start) [startstop] {Misconfigured Notebook Enabled};
\node (ssrf) [startstop, below of=start] {SSRF → Metadata Service Queried};
\node (iam) [startstop, below of=ssrf] {Temporary IAM Credentials Retrieved};
\node (api) [startstop, below of=iam] {Cosmos DB APIs Accessed};
\node (keys) [startstop, below of=api] {Primary Keys Extracted via Management Endpoint};
\node (control) [startstop, below of=keys] {Full Admin Access to Encrypted Data};

\draw [arrow] (start) -- (ssrf);
\draw [arrow] (ssrf) -- (iam);
\draw [arrow] (iam) -- (api);
\draw [arrow] (api) -- (keys);
\draw [arrow] (keys) -- (control);
\end{tikzpicture}
\caption{Attack Path in ChaosDB: From SSRF to Full Key Compromise}
\end{figure}

\subsection{Industry-Wide Trends in Cloud Security Failures}

Multiple industry reports underscore the growing mismatch between cloud complexity and organizational readiness to secure it. According to Sonatype’s 2021 Cloud Security report, 83\% of respondents believed their organizations were at serious risk of a breach due to cloud misconfiguration, and 36\% had already experienced a serious leak or breach within the past year \cite{sonatype2021}. IAM misconfiguration was cited as the most common cloud security failure, followed closely by insecure object storage permissions and disabled encryption settings.

The 2021 Qualys Cloud Security Report reinforces this narrative, with 64\% of cybersecurity professionals citing data loss or leakage as their top concern, and 46\% citing accidental exposure of credentials \cite{qualys2021}. Visibility gaps, inadequate tooling, and lack of trained personnel were identified as the main barriers to improving cloud security posture.

Palo Alto Networks’ 2024 State of Cloud-Native Security report expands the scope further, showing that 71\% of organizations faced breaches due to rushed deployments, 91\% struggle with tool sprawl, and 61\% expressed concern about AI-powered threats targeting cloud workloads \cite{paloalto2024}. These statistics point to the systemic inability of organizations to manage privilege boundaries, validate infrastructure configurations, and enforce consistent policy controls across cloud platforms.

The 2025 Checkpoint report further confirms these systemic risks, revealing that 61\% of the organizations surveyed experienced a cloud security incident in the past year and in 21\% of the cases attackers gained unauthorized access to sensitive data, underscoring the real-world consequences of misconfigured interfaces and lack of access controls \cite{checkpoint2025}. The study also identified API misuse and insufficient runtime visibility as key vectors for credential abuse and lateral movement across cloud workloads.

Across all reports, one common conclusion emerges: even robust cryptographic infrastructure such as HSMs and TPMs cannot prevent data compromise if layered on top of misconfigured, poorly governed, or overly complex cloud environments. The combination of insecure interfaces, multi-cloud fragmentation, and lack of skilled personnel makes cryptographic key management brittle and increasingly difficult to secure at scale.

\begin{table}[h]
\centering
\caption{Common Cloud-Based HSM/TPM Failure Modes}
\vspace{4pt}
\renewcommand{\arraystretch}{1.3} 
\begin{tabular}{|p{3.5cm}|p{9cm}|}
\hline
\textbf{Failure Category} & \textbf{Impact on Cryptographic Security} \\ \hline
Misconfigurations (IAM, firewall, storage) & Can bypass key isolation, allowing unauthorized access to encrypted data. \\ \hline
API Exploits & Remote attackers may invoke sensitive operations, such as key export or deletion. \\ \hline
Privilege Escalation & Grants admin-level access to cryptographic modules and key material. \\ \hline
Multi-Tenancy Risks & Potential for cross-tenant data leakage or inference of cryptographic assets. \\ \hline
\end{tabular}
\end{table}

\section{Comparative Analysis: HSMs vs. TPMs in Cloud Security}

\subsection{HSMs: Strong Hardware, Soft Targets}

Hardware Security Modules (HSMs) serve as specialized hardware-backed vaults for cryptographic key storage and operations. Cloud implementations such as AWS CloudHSM, Azure Key Vault (HSM-backed), and Google Cloud KMS rely on these modules to enforce secure encryption, decryption, and key lifecycle management.

While HSMs offer high assurance at the hardware level, their security guarantees can be compromised when integrated into complex cloud environments. In particular, API misuse, leaked credentials, and insecure development pipelines present attackers with indirect paths to compromise cryptographic workflows.

\textbf{Supply Chain and API Exploits:}  
A Wiz Security report highlights real-world cases where attackers exploited exposed CI/CD credentials and secrets—found in environment variables, build artifacts, or \texttt{.bash\_history} files—to impersonate legitimate workloads and access HSM-backed interfaces \cite{wiz2023}. These techniques bypass cryptographic enforcement not by breaking the HSM, but by abusing its trusted API surface.

\textbf{Case in Point – Google’s Internal KMS:}  
In contrast, Google’s internal KMS, as discussed in their “Secrets at Planet Scale” talk, emphasizes strong separation of duties, envelope encryption, region-level isolation, and transparent auditability \cite{googlekms2019}. This underscores the need to pair HSM use with rigorous operational controls, rather than relying on hardware guarantees alone.

\subsection{TPMs: Root of Trust, but Weak in the Cloud}

Trusted Platform Modules (TPMs) provide hardware-rooted cryptographic assurances such as attestation, measured boot, and disk encryption. In cloud environments, these are often deployed as virtual TPMs (vTPMs) attached to virtual machines or confidential compute instances.

\textbf{Hypervisor-Level Threats:}  
The “Heckler” study demonstrated that malicious hypervisors could inject crafted interrupts into VMs running under AMD SEV-SNP or Intel TDX, breaking the isolation guarantees of trusted execution environments and thereby compromising vTPM-protected workloads \cite{heckler2024}. While the physical TPM chip might remain secure, the virtual instance depending on hypervisor trust is exposed.

\textbf{vTPM Implementation Risks:}  
Earlier work on SvTPM also identified that software-emulated TPMs face significant isolation and performance issues without the support of strong TEEs \cite{svtpm2019}. This can lead to data leakage, integrity failures, or improper access to attestation records—effectively eroding the core benefits of TPM-backed trust.

\subsection{The Real Problem: Ecosystem Failures}

This comparative analysis shows that while HSMs and TPMs offer robust cryptographic foundations, their effectiveness in cloud environments is compromised by surrounding ecosystem vulnerabilities.

Attackers do not need to break encryption or tamper with secure chips. Instead, they exploit poorly scoped IAM permissions, insecure APIs, or compromised hypervisors. These indirect paths to key compromise render hardware protections ineffective unless paired with operational hardening, environment isolation, and consistent auditability.

\vspace{-4pt}

\begin{table}[H]
\centering
\caption{HSM vs. TPM/vTPM Security Comparison}
\vspace{4pt}
\label{tab:comparison}
\renewcommand{\arraystretch}{1.3}
\small
\begin{tabular}{|p{3.5cm}|p{5.5cm}|p{5.5cm}|}
\hline
\textbf{Feature} & \textbf{HSMs (Cloud)} & \textbf{TPMs / vTPMs} \\
\hline
Core Strength & Dedicated hardware for secure key management & Hardware-rooted attestation, disk encryption, and secure boot \\
\hline
Cloud Vulnerabilities & API abuse, CI/CD token leakage, supply chain compromise & Hypervisor privilege escalation, weak vTPM isolation, virtualization risks \\
\hline
Real-World Incidents & Secrets reused to access HSM APIs \cite{wiz2023} & Hypervisor injection breaks vTPM trust \cite{heckler2024} \\
\hline
Operational Challenges & Requires complex access control and continuous monitoring & Depends on TEE/hypervisor integrity; often difficult to audit \\
\hline
Key Takeaway & Hardware-secure, but APIs and operational practices are weak links & Theoretically sound, but cloud abstraction layers expand the attack surface \\
\hline
\end{tabular}
\end{table}

\section{Future Alternatives to HSMs and TPMs in Cloud Security}

While HSMs and TPMs offer hardware-backed cryptographic assurances, their effectiveness in cloud environments is increasingly compromised by the very infrastructure meant to support them. Their security boundaries are tightly scoped to physical or hypervisor-level trust, but cloud-native threats emerge at the API, orchestration, and multi-tenancy layers. This section explores evolving cryptographic approaches that aim to augment or, in some cases, challenge the current reliance on traditional HSMs and TPMs.

\subsection{Confidential Computing: Isolated Execution at Scale}
Confidential computing provides\cite{lang2022mole} hardware-enforced memory and execution isolation through trusted execution environments (TEEs). Implementations such as Intel SGX, AMD SEV-SNP, and cloud-based confidential VMs from Azure and Google enable workloads to run in isolated memory regions, shielding secrets even from privileged system software.

Azure Confidential VMs \cite{azure2024confcomp}, for example, leverage AMD SEV-SNP to ensure that memory pages used by a guest VM are encrypted and protected from access by the hypervisor. Microsoft has reported rising adoption of such technology for protecting cryptographic operations, particularly in use-cases like secure key lifecycle management and AI model inference.\cite{azure2025skr}

Despite its promise, confidential computing is not immune to side-channel threats\cite{sardar2023confidential}. Power analysis attacks, such as those demonstrated in the PLATYPUS attack framework\cite{platypus2020}, can still leak enclave-protected secrets by exploiting microarchitectural side effects. Moreover, TEEs rely on a trusted computing base (TCB) that, if compromised or misconfigured, undermines the enclave's isolation guarantees.

\subsection{Post-Quantum Cryptography: Building Crypto for the Next Era}
With quantum computing advancing, classical cryptographic primitives like RSA and ECC,commonly protected within HSMs and TPMs,face obsolescence. Post-Quantum Cryptography (PQC) aims to prepare for this threat by introducing quantum-resistant algorithms.

The NIST PQC project concluded its third round of standardization in 2024, selecting CRYSTALS-Kyber for encryption and CRYSTALS-Dilithium for signatures. These algorithms are now being tested in cloud environments by AWS, Google, and Microsoft, often in parallel with legacy cryptographic systems.

Integrating PQC into HSMs and TPMs, however, presents challenges. Existing hardware may not support large key sizes or different computational patterns required by PQC schemes. Hardware refresh cycles and firmware updates are required, and without them, cloud hardware security will lag behind cryptographic innovation.

\subsection{Decentralized and Multi-Party Key Management}
Traditional HSMs centralize key storage, creating single points of failure. Decentralized key management frameworks, including multi-party computation (MPC)\cite{blockdaemonmpc2023} and Shamir's Secret Sharing, aim to distribute trust across multiple nodes or entities.

Fireblocks and other enterprise MPC platforms already use these techniques to secure digital assets and private keys at scale. With MPC, no single node ever holds the complete key; instead, operations like signing or decryption are executed through distributed consensus, increasing resistance to breach or insider compromise.

However, MPC systems introduce coordination complexity and new attack surfaces, especially in environments lacking strong identity guarantees and synchronization mechanisms. They are best suited for use-cases where security benefits outweigh operational friction, such as digital custody and inter-organizational trust models.

\subsection{Securing vTPMs: Enhancing Trust in Virtualized Hardware}
The rise of confidential VMs and virtualized trust modules requires reevaluating TPM security in cloud-native contexts. Standard vTPMs often rely on hypervisor guarantees, which can be subverted by side-channels or control plane manipulation.

The SvTPM framework, introduced in 2019, encapsulates virtual TPM functionality within TEEs to mitigate hypervisor-level risks. Recent evaluations show that wrapping vTPMs in enclaves like Intel SGX\cite{lang2022mole} or AMD SEV improves confidentiality and operational integrity.

Still, virtualization overhead and enclave lifecycle complexity can hinder widespread adoption. Secure deployment of vTPMs must address both hardware-level isolation and orchestration-layer security policies.

\subsection{Summary and Hybrid Approaches}
While none of these approaches fully replace HSMs or TPMs yet, they offer pathways to augmenting traditional trust models in the cloud. Confidential computing delivers isolation; PQC provides cryptographic resilience; MPC removes single points of compromise; and secure vTPMs evolve the TPM paradigm for the virtual era.

Organizations are likely to adopt hybrid models that layer these techniques, balancing the strengths of hardware security with flexible, cloud-native protection strategies. As threats shift from silicon to software and orchestration, cloud cryptographic security must follow suit.

\begin{table}[H]
\centering
\caption{Compact Comparison of Emerging Alternatives}
\vspace{4pt}
\label{tab:compact-alternatives}
\renewcommand{\arraystretch}{1.3}
\small
\begin{tabular}{|p{3.5cm}|p{5.5cm}|p{5.5cm}|}
\hline
\textbf{Approach} & \textbf{Strengths} & \textbf{Limitations} \\
\hline
Confidential Computing & Isolates data-in-use using trusted hardware enclaves & Susceptible to side-channel attacks; large trusted computing base (TCB) \\
\hline
Post-Quantum Cryptography & Resistant to quantum attacks; supported by NIST standardization & Requires new algorithms, hardware support; larger keys and computational overhead \\
\hline
MPC / Decentralized Key Management & Eliminates single point of failure; enables collaborative trust models & High coordination overhead; complex implementation and maintenance \\
\hline
Secure vTPMs & Strengthens vTPM trust using enclaves like Intel SGX or AMD SEV & Enclave setup overhead; requires platform compatibility and tuning \\
\hline
\end{tabular}
\end{table}

\begin{figure}[H]
\centering
\begin{tikzpicture}[
  node distance=1.8cm,
  every node/.style={align=center, font=\small}, 
  block/.style={rectangle, rounded corners, minimum width=6cm, minimum height=1.2cm, text centered, draw=black, fill=blue!15},
  arrow/.style={thick, ->, >=stealth}
]

\node (hardware) [block] {Root of Trust:\\ TPM / SvTPM};
\node (runtime) [block, above of=hardware] {Confidential Runtime:\\ Intel SGX / AMD SEV-SNP};
\node (mpc) [block, above of=runtime] {Distributed Key Operations:\\ MPC (e.g., Fireblocks)};
\node (pqc) [block, above of=mpc] {Quantum-Resistant Crypto\cite{nistpqc2024}:\\ Kyber / Dilithium};
\node (api) [block, above of=pqc] {Cloud API Layer:\\ IAM, Secure Access Control, Auditing};

\draw [arrow] (hardware) -- (runtime);
\draw [arrow] (runtime) -- (mpc);
\draw [arrow] (mpc) -- (pqc);
\draw [arrow] (pqc) -- (api);

\end{tikzpicture}
\caption{Hybrid Cloud Cryptographic Architecture Integrating Confidential Computing, PQC, MPC, and TPM}
\label{fig:hybrid-crypto-architecture}
\end{figure}
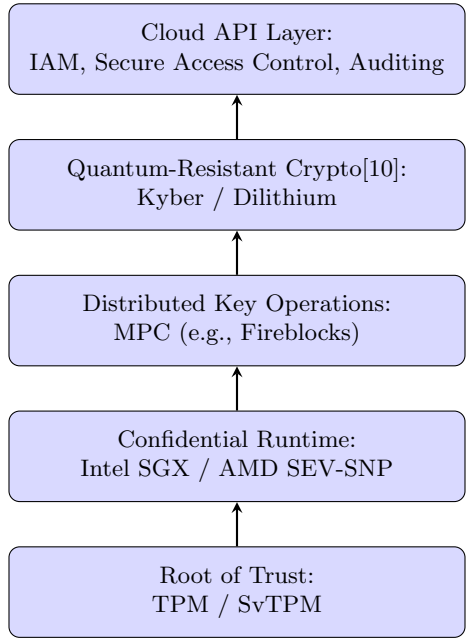

To contextualize the integration of these technologies, Figure~\ref{fig:hybrid-crypto-architecture} illustrates a layered hybrid architecture for cryptographic operations in the cloud. At the foundation lies the root of trust \cite{svtpm2019}, implemented using TPMs or secure virtual TPMs (SvTPMs), which verify system integrity at boot. Above this layer, confidential computing environments such as Intel SGX or AMD SEV-SNP isolate data-in-use during execution.

Distributed key operations\cite{fireblocksmpc2023} are handled via multi-party computation (MPC), ensuring no single device holds full key material at any time. Post-quantum cryptographic primitives, including Kyber and Dilithium, operate on top to secure data-in-transit and at-rest against quantum threats. Finally, the cloud API layer governs access control, IAM policies, and audit logging, enforcing application-layer boundaries and providing visibility into all cryptographic operations.

This architecture emphasizes modular integration, where each layer mitigates distinct threat vectors while collectively reinforcing the overall cryptographic trust chain.

This design supports a vertical trust model where:
\begin{itemize}
  \item The vTPM validates the system’s boot integrity.
  \item Confidential VMs protect runtime key material.
  \item MPC ensures distributed signing and decryption without centralized failure points.
  \item PQC algorithms future-proof the cryptographic layer.
  \item IAM/API gateways limit access and log every operation.
\end{itemize}

\section{Conclusion and Recommendations}

This research demonstrates that while HSMs and TPMs offer strong hardware-based assurances, their effectiveness in cloud environments is repeatedly compromised by surrounding ecosystem weaknesses. Misconfigured APIs, exposed credentials, compromised hypervisors, and lack of isolation mechanisms allow attackers to bypass hardware protections without needing to break cryptographic primitives.

\subsection*{Key Observations}

\begin{itemize}
    \item Cloud-hosted HSMs are undermined by insecure API surfaces, token reuse, and supply-chain leakage \cite{wiz2023}.
    
    \item Virtual TPMs (vTPMs), while offering hardware-rooted trust, are vulnerable to control-plane attacks such as interrupt injection and VM escape, as demonstrated by the ``Heckler'' exploit \cite{heckler2024}.
    
    \item Wrapping vTPM logic inside trusted execution environments (e.g., SvTPM over SGX) significantly enhances confidentiality and mitigates hypervisor risk \cite{svtpm2019}.
\end{itemize}

\subsection*{Recommendations for Cloud-Native Cryptographic Security}

\begin{itemize}
    \item \textbf{Adopt confidential computing}: Deploy AMD SEV-SNP or Intel TDX-based confidential VMs to isolate memory during runtime, even from privileged host layers \cite{azure2024confcomp}.
    
    \item \textbf{Protect vTPMs with enclaves}: Use enclave-based implementations such as SvTPM to secure virtual trust anchors against host manipulation.
    
    \item \textbf{Prepare for post-quantum cryptography}: Begin transitioning to NIST-standardized PQC algorithms like CRYSTALS-Kyber and Dilithium within HSM and vTPM infrastructure \cite{nistpqc2024}.
    
    \item \textbf{Adopt decentralized key models}: Use multi-party computation (MPC) and secret sharing to eliminate single points of cryptographic failure \cite{fireblocksmpc2023}.
    
    \item \textbf{Harden API surfaces}: Enforce least-privilege IAM, rotate tokens frequently, restrict credential scope, and instrument real-time auditing for all key-related operations.
\end{itemize}

Future-proofing cryptographic security in the cloud requires shifting focus from hardware guarantees to holistic system architecture. Hardware-secure modules must operate within environments that enforce strict privilege, isolation, and lifecycle control. Only then can HSMs and TPMs meet their original security objectives under modern threat models.

\clearpage

\bibliographystyle{ieeetr}

\end{document}